\documentstyle[aps,prd,amssymb,12pt,preprint,tighten,epsf]{revtex}
\begin{document}
\draft
\title{Gravitational Instability of a Kink}
\author{W. Barreto${}^{1,2}$,
	R. G\'{o}mez${}^{1}$,
	L. Lehner${}^{1}$ and
	J. Winicour${}^{1}$}
\address{
${}^{1}$ Department of Physics and Astronomy,
 	 University of Pittsburgh,
 	 Pittsburgh, PA 15260 \\
${}^{2}$ Laboratorio de F\'{\i}sica Te\'orica,
	Departamento de F\'{\i}sica,
	Escuela de Ciencias, N\'ucleo de Sucre
	Universidad de Oriente,
	Cuman\'a, Venezuela}
\maketitle

\begin{abstract}

We study the equilibria of a self-gravitating scalar field in the
region outside a reflecting barrier. By introducing a potential
difference between the barrier and infinity, we create a kink which
cannot decay to a zero energy state. In the realm of small amplitude,
the kink decays to a known static solution of the Einstein-Klein-Gordon
equation. However, for larger kinks the static equilibria are
degenerate, forming a system with two energy levels. The upper level is
unstable and, under small perturbations, decays to the lower energy
stable equilibrium. Under large perturbations, the unstable upper level
undergoes collapse to a black hole.  The equilibrium of the system
provides a remarkably simple and beautiful illustration of a turning
point instability.

\end{abstract}


\section{Introduction}

We describe here a simple model of a system exhibiting a
gravitationally induced transition from an unstable static equilibrium
to a lower energy stable equilibrium (for small perturbations) or to a
black hole (for large perturbations).  The system is the general
relativistic version of a spherically symmetric scalar wave $\Phi$
reflecting off an inner mirror of radius $R$. By holding the mirror at
a fixed potential with respect to infinity, we create kink boundary
conditions which do not allow decay of the field to a zero energy
state. In the nongravitating case, any initial configuration would
decay to the static solution of the wave equation, $\Phi = AR/r$, where
$A$ is the potential of the mirror; and this static solution is stable
under perturbations. In the self-gravitating case we will show there is
a critical value of kink potential $A_c$ about which the system
displays a classic example of a turning point
instability~\cite{ipser,sork}, responsible for the above phenomena.
Above this critical potential, there are no equilibria and all
configurations collapse to a black hole.  The behavior of this system
illustrates the remarkable richness of the physical properties that
result from the mixture of long range gravitational forces and
nonlinearity.

Our theoretical model is the null-timelike initial value problem for a
spherically symmetric space-time satisfying the coupled
Einstein-Klein-Gordon equations for a massless scalar field.  We pose
this mixed boundary value problem in the region of space-time outside a
timelike inner boundary and to the future of an initial null
hypersurface. The mathematical details of this problem have been
described before in an investigation of the decay of scalar wave
tails~\cite{schmidttail}. In that case, the scalar field was set to zero
on the inner boundary, so that the boundary acted as a perfectly
reflecting mirror. Here, by holding the inner boundary at a fixed
potential, its role as a mirror is unchanged due to the local gauge
invariance of the system with respect to the transformation
$\Phi\rightarrow \Phi+constant$. However, the potential difference with
respect to infinity introduces nontrivial global effects.

We set the geometry of the mirror to correspond to a surface of radius
$R$ enclosing a flat region of spacetime.  The evolution of the system
then depends uniquely upon the characteristic data for the scalar field
on the initial null hypersurface, which extends from the mirror to
future null infinity ${\cal I}^+$.

We describe this system in terms of a null coordinate system with the
spherically symmetric line element
\begin{equation}
    ds^2= e^{2\beta}du({V \over r}du+2dr)
    - r^2(d\theta^2 +\sin^2\theta d\phi^2). \label{eq:metric}
\end{equation}
In these coordinates, the Einstein-Klein-Gordon equations reduce
to~\cite{EKG,X1986}
\begin{equation}
     \beta_{,r}= 2\pi r(\Phi_{,r})^2    \label{eq:beta}
\end{equation}
\begin{equation}
     V_{,r}=e^{2\beta}             \label{eq:V}
\end{equation}
and the scalar wave equation $\Box \Phi = 0$, which takes the form
\begin{equation}
     2(r \Phi)_{,u r} = r^{-1}(rV\Phi_{,r})_{,r} . \label{eq:SWE}
\end{equation}

The initial null data necessary for evolution consists of
$\Phi(u_0,r)$, $r\ge R$, at initial time $u_0$. (We take $u_0 =0$). At
the mirror, we set $\Phi(u,R)=A=constant$, with the gauge condition
that $\Phi(u,\infty)=0$. We adopt the coordinate condition
$\beta(u,R)=0$.  The condition that the metric match continuously to a
flat interior for $r<R$ requires $V(u,R)=R$. (The normal derivatives of
$\Phi$ and the metric do not match continuously across $r=R$ in accord
with the sheet stresses associated with the mirror).

With these conditions the scalar field and metric components have a
unique future evolution. The resulting metric does not have an asymptotic
Minkowski form at ${\cal I}^+$.  This is characterized by the quantity
$H(u)=\beta (u,\infty)$ which relates  Bondi time $t$ at ${\cal I}^+$
to the proper time $\tau=u$ at the reflecting boundary according to
$dt/d\tau = e^{2H}$. Bondi time is the physically relevant time for
distant observers.  The Bondi mass of the system can be expressed in
either an asymptotic or integral form~\cite{waveforms}:
\begin{eqnarray}
  M(u)&=&{1\over 2}e^{-2H}r^2({V\over r})_{,r}\,\Bigg| _{r=\infty}\nonumber \\
      &=& 2\pi\int_R^\infty e^{2(\beta-H)} r^2 (\Phi_{,r})^2 dr .
    \label{eq:mint}
\end{eqnarray}
(Note that the mass content of a null hypersurface only depends upon field
values in the hypersurface and is independent of their time
derivative).

In the flat space limit, (\ref{eq:beta}) and (\ref{eq:V}) imply $\beta
=0$ and $V=r$ and (\ref{eq:SWE}) reduces to the wave equation in
null-radial coordinates:
\begin{equation}
    \Box^{(2)} g :=  2g_{,ur} -g_{,rr}=0 , \label{eq:fwe}
\end{equation}
where $g=r\Phi$. The general solution satisfying the boundary condition for a
kink potential $A$ is
\begin{equation}
       g(u,r)=AR+f(u/2+r)-f(u/2+R).  \label{eq:flatg}
\end{equation}
The first term describes a static monopole; the second, an incoming
wave; and the third, its reflection off the mirror. At late retarded
time, the solution asymptotically approaches the static equilibrium state
$\Phi\rightarrow AR/r$. This state is also the solution of the
variational problem for the energy, $\delta M=0$ subject to the
constraint that $\delta \Phi$ vanishes at the mirror and at infinity. In
this flat space example,
\begin{equation}
 M= 2\pi\int_R^\infty (\Phi_{,r})^2 r^2  dr
\end{equation}
and its variation is given by
\begin{equation}
 \delta M=-4\pi\int_R^\infty [\Phi_{,r} r^2]_{,r} \delta \Phi dr.
\end{equation}
In the flat space case, the static solution $\Phi=A/r$ is the unique
extremum satisfying the boundary conditions (which rule out the trivial
solution $\Phi=constant$). Furthermore, the second variation of the
mass about this state is positive, so that it represents a stable
equilibrium, which the system approaches after radiating its excess
energy to infinity.

In Sec.\ref{sec:energ}, we show that for small kink potentials these
features extend to the curved space, self-gravitating case. But, above
the critical kink potential, equilibria do not exist. We show that this
critical potential marks a turning point instability about which
the equilibrium configurations bifurcate into stable and unstable form.
In Sec.\ref{sec:collap}, we present numerical simulations of the
evolution of stable and unstable kinks.

\section{Energetics of self-gravitating kinks} \label{sec:energ}

At the initial retarded time, the null data $\Phi(0,r)$ and the
constrained kink potential $A=\Phi(u,R)$ uniquely determine the future
evolution of the system. In order to discuss the energetics of these
states it is convenient to describe the corresponding configuration
space in terms of $\Phi(V)=\Phi(0,r(V))$, where $V$ is considered the
independent variable. Here $V$, as determined by (\ref{eq:beta}) and
(\ref{eq:V}), is the affine parameter along the outgoing null cone. The
affine freedom is fixed by the boundary conditions
$V(R)=r(R)=R$ and $r^{\prime}(R)= 1$, where we denote $\partial_V
f=f^{\prime}$. Then (\ref{eq:beta}) and (\ref{eq:V}) can be reexpressed
as $r=f$, where $f$ satisfies
\begin{equation}
   f^{\prime \prime} = -4\pi (\Phi^{\prime})^2 f, \label{eq:r}
\end{equation}
which determines $r(V)$ uniquely in terms of the initial data. The
mass of of the state $\Phi(V)$ can be obtained by inserting the
solution $r(V)$ of (\ref{eq:r}) into the first version of
(\ref{eq:mint}), which gives
\begin{equation}
     M={1\over 2}(r-Vr^{\prime})\Bigg| _{\infty},
   \label{eq:asymm}
\end{equation}
when reexpressed with $V$ as the independent variable.

\subsection{Static equilibria} \label{sec:equil}

We now show that the asymptotically flat static solution $\Psi$ of the
Einstein-Klein-Gordon system is an extremum of the energy, subject to a
fixed kink potential.  This solution~\cite{jnw1968}, which is the
analog of the $1/r$ solution in a Minkowski background, can be obtained
in null coordinates by setting $\Phi_{,u}=0$ in the wave
equation~(\ref{eq:SWE}). This gives
\begin{equation}
	  rV \Psi_{,r} = const,  \label{eq:lstat}
\end{equation}
whose solution, after using (\ref{eq:beta}) and (\ref{eq:V}) to
eliminate the $r$-dependence, is
\begin{equation}
        \Psi(V) = {1 \over {4\sqrt{\pi} \>{\cosh} \alpha}} \ln
        \bigl [{{V + R\> (e^{2\>\alpha} - 1)} \over
         {V + R\> (e^{-2\>\alpha} - 1)}} \bigl ] ,\label{eq:static}
\end{equation}
with $r(V)=r_\Psi$ given by
\begin{equation}
        r_\Psi^2 =  e^{-4\> \alpha \tanh \alpha}
               [V + R\> (e^{-2\>\alpha} - 1)]^{1\> -\> \tanh \alpha}
               [V + R\> (e^{2\>\alpha} - 1) ]^{1\> +\> \tanh \alpha}.
                \label{eq:rstatic}
\end{equation}
Here the integration constant $\alpha$ determines
the kink potential,
\begin{equation}
       A_\Psi(\alpha)={\alpha \over \sqrt{\pi}\cosh\alpha}. \label{eq:kink}
\end{equation}
The spacetime has a naked singularity when analytically extended to
$r=0$~\cite{jnw1968}. The Bondi mass of this solution is
\begin{equation}
    M_{\Psi}(\alpha)=2R\sinh^2\alpha e^{-2\alpha\tanh\alpha}.
             \label{eq:statm}
\end{equation}

In order to investigate the behavior of the Bondi mass with respect to
variations about $\Psi$, we consider the solutions of the second order
differential equation (\ref{eq:r}) for
an arbitrary configuration $\Phi$.
The solution $r$ is uniquely determined by the boundary conditions
$r(R)=R$ and $r^{\prime}(R)=1$ and has asymptotic behavior $r\sim
e^{-2H}V+2M+O(1/V)$ as $V\rightarrow \infty$. We define a second
independent solution $\tau$ by requiring asymptotic behavior
$\tau\sim V+O(1/V)$. For a general configuration $\Phi$, this
normalizes the Wronskian of these solutions to the Bondi mass,
\begin{equation}
  2M =  r\tau^\prime- \tau r^\prime. \label{eq:wronsk}
\end{equation}

Now, for a fixed kink potential, consider the $\epsilon$-dependent
family of perturbed configurations $\Phi = \Psi +\epsilon \phi$,
subject to the boundary condition $\phi(R)= \phi(\infty)=0$. For
$\epsilon=0$, we have $r=r_\Psi$ and $\tau=\tau_\Psi$, with $r_\Psi$
given by (\ref{eq:rstatic}) and
\begin{equation}
 \tau_\Psi^2=[V + R\> (e^{-2\>\alpha} - 1)]^{1\> +\> \tanh \alpha}
            [V + R\> (e^{2\>\alpha} - 1) ]^{1\> -\> \tanh \alpha}.
\end{equation}
To first order in $\epsilon$, $\delta r =r-r_\Psi$ satisfies the
perturbed version of (\ref{eq:r}):
\begin{equation}
  \delta r^{\prime\prime}=-4\pi \delta r(\Psi^{\prime})^2
        -8\pi\epsilon r_\Psi\Psi^{\prime}\phi^{\prime} . \label{eq:pert}
\end{equation}
This linear differential equation for $\delta r$ is identical to
(\ref{eq:r}) except for a inhomogeneous term. Its solution can therefore
be expressed in terms of the two independent homogeneous solutions
$r_\Psi$ and $\tau_\Psi$ as~\cite{codd}
\begin{equation}
 \delta r=-{4\pi \epsilon\over M_\Psi}\int_R^V dW
[r_\Psi(W)\tau_\Psi(V)-\tau_\Psi(W)r_\Psi(V)]
           r_\Psi(W)\Psi^{\prime}(W)\phi^{\prime}(W).
    \label{eq:deltar}
\end{equation}
Note that this satisfies the proper boundary conditions $\delta r(R)
=\delta r^{\prime}(R)=0$. By inserting this solution for $\delta r$
into (\ref{eq:asymm}), we obtain for the first variation of the Bondi mass
\begin{equation}
  M-M_\Psi= 4\pi \epsilon \int_R^\infty dV
          \tau_\Psi r_\Psi \Psi^{\prime}\phi^{\prime}.
       \label{eq:delta1}
\end{equation}
But direct calculation gives
\begin{equation}
   \tau_\Psi r_\Psi \Psi^{\prime}=-{R\sinh\alpha \over \sqrt{\pi}}
	 e^{-2\alpha\tanh\alpha} = constant
\end{equation}
so that (\ref{eq:delta1}) integrates to yield $M-
M_\Psi=O(\epsilon^2)$.

Thus the static solutions are equilibrium configurations. For
sufficiently small values of $\alpha$, the second variation of the
Bondi mass will be positive, as in the flat space case. In that regime,
small perturbations can be dissipated by mass loss due to scalar
radiation.  However, if there exists a perturbation which lowers
the Bondi mass the static equilibria cannot be stable since further
radiative mass loss would drive the system away from equilibrium. We
next see that this indeed occurs in the strongly nonlinear regime.

\subsection{The turning point instability}

The one-parameter family of static equilibria has kink potential
$A_\Psi(\alpha)$, given by  (\ref{eq:kink}), which increases
monotonically with $\alpha$ from $A_\Psi(0)=0$ until it reaches a
maximum at the turning point $\alpha_c\approx 1.199$ satisfying
$\alpha_c\tanh\alpha_c=1$.  Above $\alpha_c$, $A_\Psi(\alpha)$
monotonically decreases to $0$ as $\alpha\rightarrow\infty$. Thus,
below $A_\Psi(\alpha_c)$, there are two static equilibria for each kink
amplitude. Similarly, the mass $M_\Psi(\alpha)=0$ increases
monotonically from $M_\Psi(0)=0$ to a maximum at the same turning point
$\alpha_c$ and then decreases monotonically to the black hole limit,
$M_\Psi(\alpha)\rightarrow R/2$, as $\alpha\rightarrow\infty$.

The quantity $\kappa$, defined by $dM_\Psi/d\alpha=\kappa
dA_\Psi/d\alpha$, plays an important role in the stability of this
system. Explicitly, \begin{equation}
   \kappa =  4\sqrt{\pi}R\sinh(\alpha)e^{-2\alpha\tanh\alpha}.
\end{equation} Denoting ${\dot f}=df/d\alpha$, it is easy to verify
that ${\dot \kappa}{\ddot A}_\Psi\approx .65R >0$ for
$\alpha=\alpha_c$.

These are precisely the criteria for the application of a theorem~\cite{sork}
regarding the onset of instability at a turning point in a
one-parameter family of equilibria. We present here a finite
dimensional version of the argument behind the theorem as applied to
the present case. Proofs for both the finite dimensional
and function space cases are given by Sorkin~\cite{sork}.

Consider then the configuration space $\{\Phi(A,X^1,...,X^n)\}$ whose
elements for
each choice $(A,X^i)$ represent a function $\Phi(V;A,X^i)$ with
boundary conditions $\Phi(R;A,X^i)=A$ and $\Phi(\infty;A,X^i)=0$. Let
$\Psi(\alpha)$ represent the one parameter family of static equilibria,
regarding $(A_\Psi,X^i)$ as functions of $\alpha$. We now show that for
some $\delta \alpha >0$ all the equilibria in the range $\alpha_c
<\alpha <\alpha_c +\delta \alpha$ are unstable.

The proof is based upon the equilibrium conditions, $M_{\Psi,A}=\kappa$
and $M_{\Psi,i}=0$, where $f_{,A}=\partial_A f$ and
$f_{,i}=\partial_{X^i}f$. Differentiating these conditions with respect
to $\alpha$ along the equilibrium sequence gives
\begin{equation}
      {\dot \kappa}=M_{\Psi,AA}{\dot A}_\Psi +M_{\Psi,Ai}{\dot X}^i
\end{equation}
and
\begin{equation}
         0={\dot M}_{\Psi,i}{\dot X}^i =
       M_{\Psi,iA}{\dot A}_\Psi{\dot X}^i +\Delta,
    \label{eq:mdot}
\end{equation}
where $\Delta=M_{\Psi,ij}{\dot X}^i{\dot X}^j$ represents the second
variation of the mass with respect to the perturbation ${\dot X}^i$.
Equation (\ref{eq:mdot}) implies that $\Delta=0$ at the turning point
where ${\dot A}_\Psi=0$, , so that this mode has neutral stability.
Taking a further derivative of (\ref{eq:mdot}) and evaluating at the
turning point, we obtain
\begin{equation}
   {\dot \Delta}=-M_{\Psi,iA}{\ddot A}_\Psi{\dot X}^i
              =-{\dot \kappa}{\ddot A}_\Psi. \label{eq:deltadot}
\end{equation}
As remarked above, ${\dot \kappa}{\ddot A}_\Psi>0$ at the turning
point so that (\ref{eq:deltadot}) implies the instability of this mode
in a neighborhood $\alpha>\alpha_c$.

\subsection{Unstable modes}

Since there are no other turning points, these general considerations
suggest that the equilibria $\Psi$ are stable below $\alpha_c$ and
unstable above $\alpha_c$, although the theorem only strictly implies
the onset of instability in a neighborhood of $\alpha_c$.  In order to
provide further insight, we now consider specific behavior of
the second variation of $M_\Psi$.

For an arbitrary perturbation, $\Phi = \Psi +\epsilon \phi$, we can
follow the approach in Sec.~\ref{sec:equil} which led to the expression
(\ref{eq:delta1}) for the first variation of the mass and proceed
further to determine the second variation. At order $\epsilon^2$, this
leads to
\begin{equation}
  M-M_\Psi=2\pi\epsilon^2 \int_R ^\infty dV
   \tau_\Psi( r_\Psi \phi^{\prime}\phi^{\prime}
         +2\delta r\Psi^{\prime}\phi^{\prime}),
   \label{eq:delta2}
\end{equation}
where $\delta r$ is given by the integral (\ref{eq:deltar}). Because
$\delta r$ depends nonlocally on the perturbation $\phi$,
(\ref{eq:delta2}) does not give straightforward information regarding
the sign of the second variation. However, the first term in the
integrand is positive definite and depends locally on the
perturbation.  Therefore, for perturbations confined to a sufficiently
compact region, the first term dominates the $\delta r$ term and the
second variation is positive.  This leads us to consider large
length scale perturbations in the search for unstable modes.

One such perturbation can be obtained from the static solution by combining
a change in $\alpha$ with a translation,
\begin{equation}
  \delta\phi(V)=\Psi(V+\delta R;\alpha+\delta\alpha)-\Psi(V;\alpha),
\end{equation}
with $\delta\alpha$ and $\delta R$ adjusted to set $\delta\phi(R)=0$
so that the kink potential is held fixed. This requires
\begin{equation}
   \delta R={R(e^{-2(\alpha+\delta\alpha)}
            e^{4\alpha\cosh(\alpha+\delta\alpha)/\cosh(\alpha)}
            -e^{2(\alpha+\delta\alpha)})
           \over
            1-e^{4\alpha\cosh(\alpha+\delta\alpha)/\cosh(\alpha)}} .
\end{equation}
For this perturbation, the change in $M$ can be worked out analytically
to second order (using Maple) from (\ref{eq:wronsk}), although the
final expression is too lengthy to present here. The important point is
that the second order variation of $M$ is positive for
$\alpha<\alpha_c$, vanishes precisely for $\alpha=\alpha_c$ and turns
negative in a neighborhood $\alpha>\alpha_c$.  This corroborates our
expectation that perturbations are stable below $\alpha_c$ and go
unstable as we pass through the turning point. (It is curious that
beyond the turning point $M-M_\Psi$ again becomes positive for values
of $\alpha$ greater than approximately $2.0$).

\section{Gravitational collapse of unstable kinks} \label{sec:collap}

We now examine the behavior of this system numerically, using a null
cone evolution algorithm for nonlinear scalar waves developed in
Ref's~\cite{Gom} and~\cite{EKG}. The algorithm is based upon the
compactified radial coordinate $x=r/(R+r)$, so that ${\cal I}^+$ is
represented by a finite grid boundary, with $x=1/2$ at the mirror and
$x=1$ at ${\cal I}^+$.  The code has been tested to be globally second
order accurate, i.e.  the error in global quantities such as the Bondi
mass is $O(\Delta x^2)$ in terms of the grid spacing $\Delta x$.

In the case of a mirror boundary at zero potential (no kink), as
studied in Ref.~\cite{schmidttail}, a weak scalar field radiates
completely to infinity, so that the final mass of the system is zero.
However, above a critical strength, the scalar field undergoes
gravitational collapse to form a horizon. In this case, some of the
scalar energy is radiated to infinity and the remainder crosses the
horizon and contributes to the final black hole mass. The mirror itself
must fall into the horizon for otherwise it would continue to reflect
the scalar field until all scalar energy were radiated to infinity.
Near the critical strength, the sensitivity of the final mass is
somewhat analogous to the critical behavior studied by
Choptuik~\cite{chsouth,choptprl}, except there is now a mass gap
because the final black hole must have a mass larger than $R/2$ in
order to contain the mirror. In the present case of a mirror with
nonvanishing kink potential, we would expect this behavior to be
modified in several ways.

\begin{figure}
\centerline{\epsfxsize=4in\epsfbox{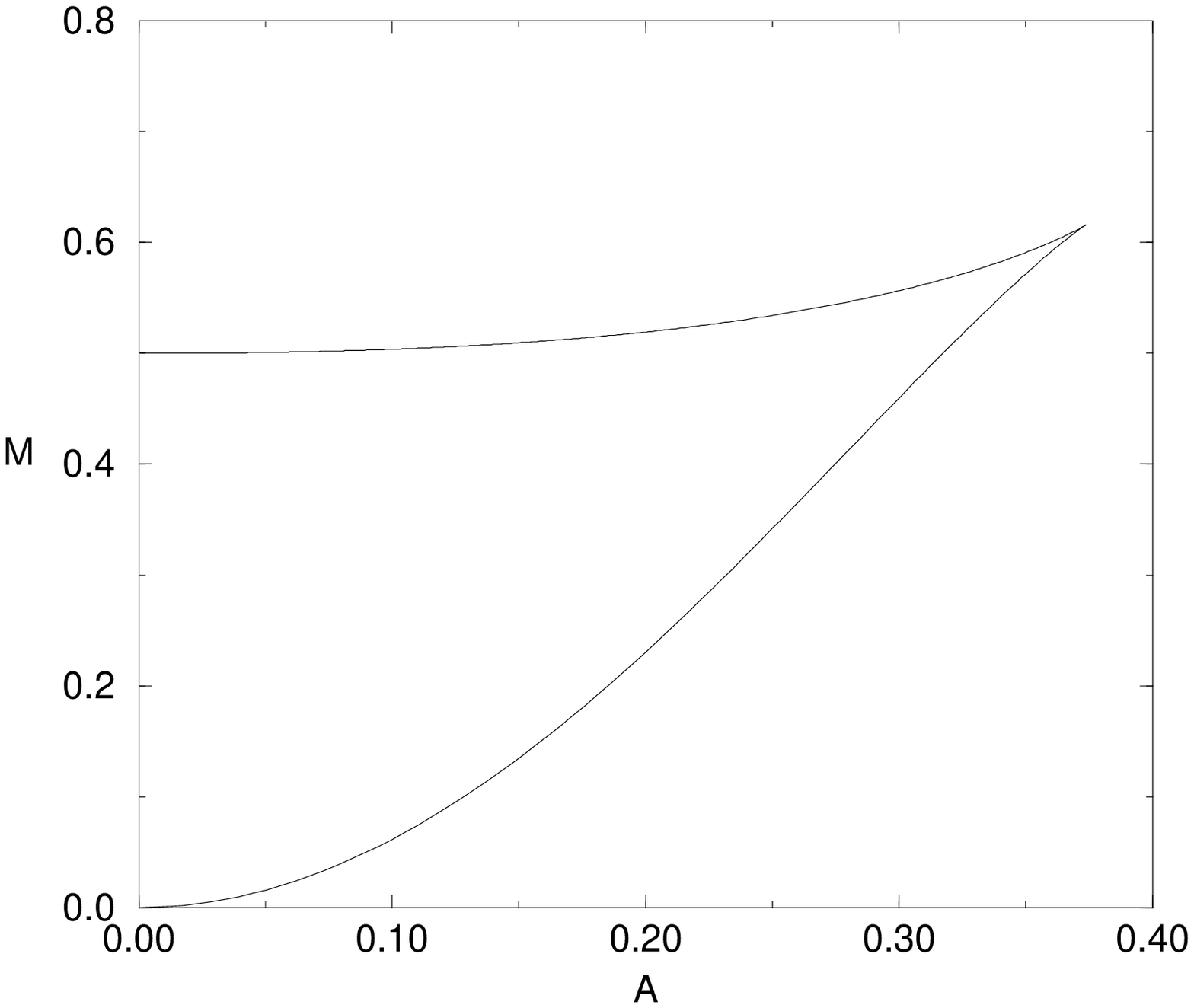}}
\caption{Bondi mass as a function of amplitude, for the static
equilibria.}
\label{fig:mass}
\end{figure}

First, we consider kink potentials $A<A_\Psi(\alpha_c)$. In this case,
two static equilibrium states $\Psi(\alpha_-)$ and $\Psi(\alpha_+)$ are
possible, $\alpha_-<\alpha_c<\alpha_+$, with the former expected to be
stable and the latter unstable. The double valued behavior of $M(A)$ is
graphed in Figure~\ref{fig:mass}. It shows that
$M_\Psi(\alpha_+)>M_\Psi(\alpha_-)$.  Consequently, the following
scenarios are expected. A state close to $\Psi(\alpha_-)$ should return
to $\Psi(\alpha_-)$ but a state close to $\Psi(\alpha_+)$ could either
evolve toward $\Psi(\alpha_-)$ or collapse to a black hole. These
scenarios have been confirmed by the following numerical simulations.
(All these simulations are run setting $R=1$).

\begin{figure}
\centerline{\epsfxsize=4in\epsfbox{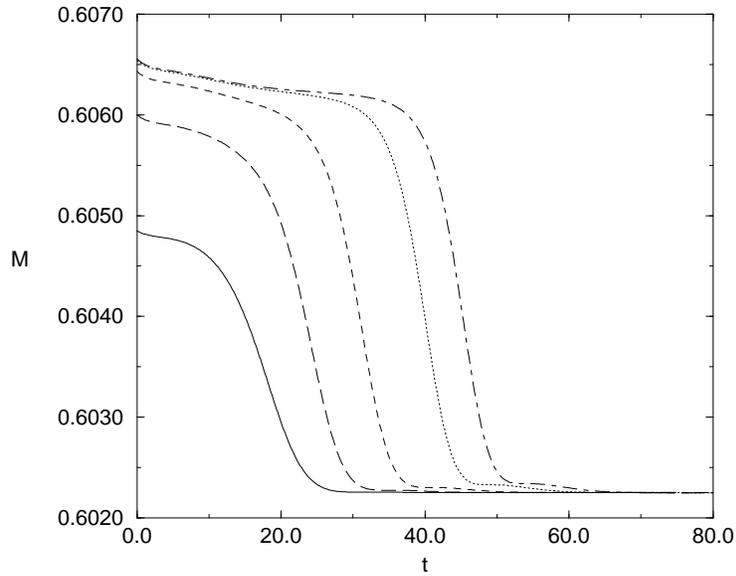}}
\caption{Bondi mass as a function of Bondi time for
perturbations of a stable equilibrium. Different curves represent the
different values of the perturbation parameter $\lambda$: $0.15$
(continuous line); $0.18$
(dashed line); $0.19$ (small--dashed line); $0.1925$ (dotted line);
$0.1928$ (dot--dashed line).  The system collapses to a black hole
for $\lambda$ greater than $\lambda_{c}\approx 0.1929$.}
\label{fig:minus}
\end{figure}

With kink potential $A=A_\Psi(\alpha_-)\approx 0.36563$ for
${\alpha_-=1}$, we study the stable equilibrium by evolving initial
data of the form $\Phi=\Psi(\alpha_-)+\lambda(1-2x)/2r$.  For values of
$\lambda$ less than a critical value $\lambda_c\approx 0.1929$, the
system returns to the equilibrium state $\Psi(\alpha_-)$ at late times.
Fig.~\ref{fig:minus} graphs the corresponding time behavior of the
Bondi mass showing its  asymptotic approach to the equilibrium value
$M\rightarrow M_\Psi(\alpha_-)\approx .60220$. For $\lambda>
\lambda_c$, the system collapses to form a black hole.

\begin{figure}
\centerline{\epsfxsize=4in\epsfbox{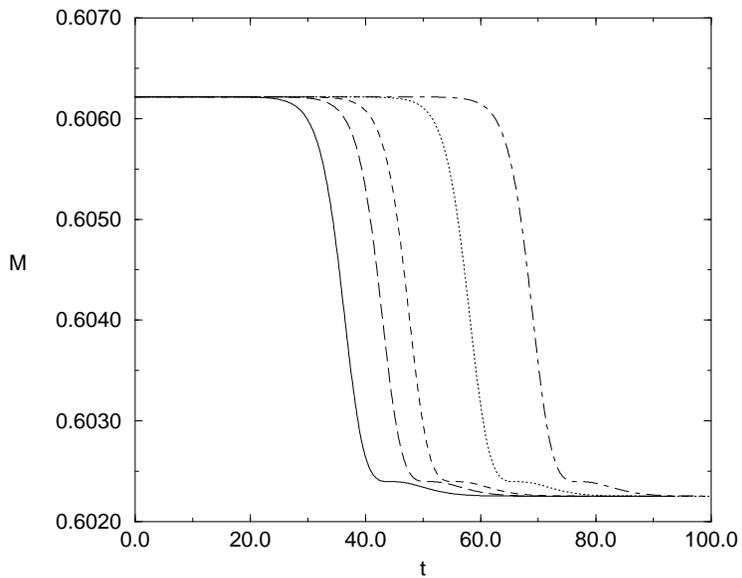}}
\caption{Bondi mass as a function of Bondi time for perturbations of an
unstable equilibrium. The different values of the perturbation
parameter $\lambda$ are: $0$ (continuous line); $10^{-4}$ (dashed
line); $(1.2)\,\,10^{-4}$ (small--dashed line);
$(1.3)\,\,10^{-4}$ (dotted line); $(1.31)\,\,10^{-4}$ (dot--dashed
line). For these values the system makes a transition to the stable
equilibrium. The system collapses to a black hole for
$\lambda_{c}\approx (1.311)\,\,10^{-4}$.}
\label{fig:plus}
\end{figure}

Next, for the same kink potential $A\approx 0.36563$, we study the
evolution of the unstable equilibrium by evolving initial data of the
form $\Phi=\Psi(\alpha_+)+\lambda(1-2x)/2r$. (Here $\alpha_+\approx
1.42429$). Fig.~\ref{fig:plus} graphs the corresponding behavior of the
Bondi mass. In this case, because of the instability of the static
equilibrium $\Psi(\alpha_+)$, the system undergoes a dynamical change,
even when $\lambda=0$ because of the perturbation introduced by
discretization error in the initial data. For small $\lambda$, the
state $\Psi(\alpha_+)$ makes a transition to the stable state
$\Psi(\alpha_-)$. But again there is a critical value $\lambda_c\approx
(1.311)\,\,10^{-4}$ above which the final state is a black hole.

\begin{figure}
 \centerline{\epsfxsize=4in\epsfbox{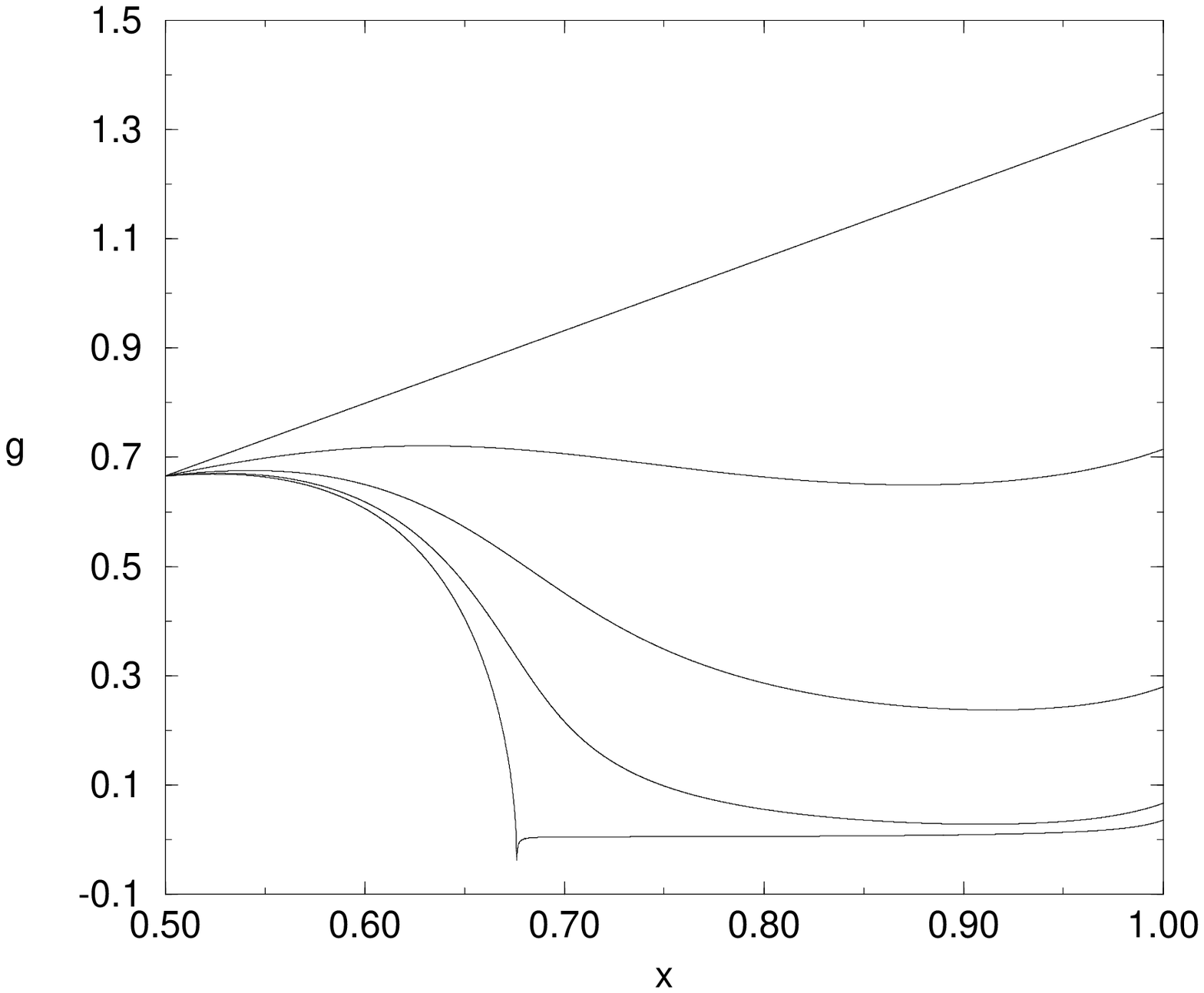}}
 \caption{Evolution of $g$. The initial data is
 $g(0,x)=2x(A_c+\lambda)$, for $\lambda=0.3$. The upper line represents
 $u=0$; and the lower one, $u=1.19$. The system collapses to a black
 hole.}
 \label{fig:large}
 \end{figure}

\begin{figure}
 \centerline{\epsfxsize=4in\epsfbox{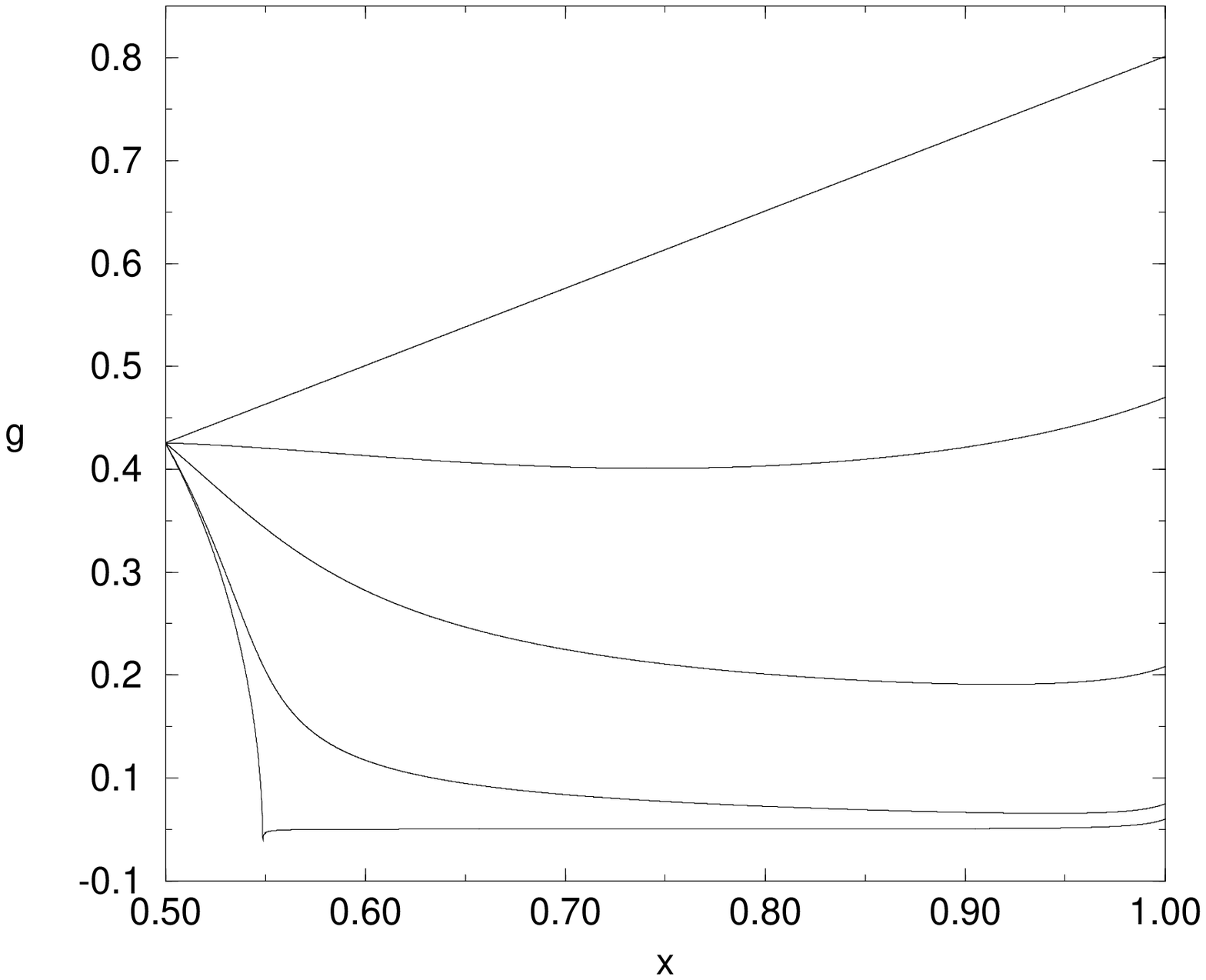}}
 \caption{Evolution of $g$. The initial data is
 $g(0,x)=2x(A_c+\lambda)$, for $\lambda=10^{-2}$. The upper line
 represents $u=0$, and the lower one $u=15.33$. The system collapses to
 a black hole.}
 \label{fig:small}
 \end{figure}

Now consider a kink potential $A>A_c$, for which no static equilibria
exist. With these boundary conditions, we would expect any initial
state to undergo collapse to a black hole. We explore this by
considering initial data of the form $\Phi=2(A_c+\lambda)/(r+R)$.
Figs. \ref{fig:large} and \ref{fig:small} show the resulting behavior
of $g$ as a function of $x$ at several times, for
$\lambda=0.3$ and $\lambda=10^{-2}$, respectively. At late times, these
states indeed form a black hole, as evidenced by the flattening of
$g$, outside the cusp being formed at the black hole radius, as the
exterior field sheds its ``hair''.

In summary, the behavior of this system is in complete accord with theoretical
expectations.

\acknowledgements

We benefited from research support from the National Science Foundation
under Grant PHY9510895 to the University of Pittsburgh and from
computer time made available through the grant PHY850023P from the
Pittsburgh Supercomputing Center. W.B. is grateful for the hospitality
shown to him by the Relativity Group of the University of Pittsburgh,
during his sabbatical year. He was supported in part by the Consejo de
Investigaci\'on de la Universidad de Oriente, Venezuela.


\end{document}